\documentclass[12pt]{article}

\usepackage{graphicx}

\def\fun#1#2{\lower3.6pt\vbox{\baselineskip0pt\lineskip.9pt
  \ialign{$\mathsurround=0pt#1\hfil##\hfil$\crcr#2\crcr\sim\crcr}}}
\def\lap{\mathrel{\mathpalette\fun <}}
\def\gap{\mathrel{\mathpalette\fun >}}

\title{Tracing Black Hole Mergers Through Radio Lobe Morphology} 

\author
{David Merritt,$^{1}$ R. D. Ekers$^{2,3}$\\
\\
\normalsize{$^{1}$Department of Physics and Astronomy, Rutgers University,}\\
\normalsize{New Brunswick, NJ, 08903, USA}\\
\normalsize{$^{2}$Australia Telescope National Facility, CSIRO,}\\
\normalsize{P. O. Box 76, Epping, NSW 2121, Australia}\\
\normalsize{$^3$Radio Astronomy Laboratory, 623 Campbell Hall,}\\
\normalsize{University of California, Berkeley, CA 94720}\\
}

\date{}

\begin{document} 

\baselineskip16pt

\maketitle 

\begin{abstract}

Binary supermassive black holes are produced by galactic
mergers as the black holes from the two galaxies fall to the
center of the merged system and form a bound pair.
The two black holes will eventually coalesce in an enormous burst
of gravitational radiation.
Here we show that the orientation of a 
black hole's spin axis would change dramatically even in
a minor merger, leading to a sudden flip in the direction
of any associated jet.
We identify the winged or X-type radio sources with galaxies in which 
this has occurred. The implied coalescence rate is similar 
to the overall galaxy merger rate, suggesting that the prospects are
good for observing gravitational waves from coalescing supermassive
black holes.

\end{abstract}

\clearpage

The detection of gravitational radiation from coalescing 
supermassive black holes (SBHs) would 
constitute a rigorous test of general relativity
in the strong-field limit ({\it 1}).
The expected event rate is uncertain however, 
because the emission of gravitational waves is negligible until the 
separation between the SBHs falls below $\sim 10^{-3}-10^{-2}$ pc.
By contrast, simulations of binary SBHs at the centers of
galaxies suggest that binary decay may stall at separations 
of $\sim 1$ pc, too great for the efficient emission of gravitational 
waves ({\it 2}).
It is unclear whether stellar- or gas-dynamical processes are capable
of bridging this gap in a time shorter than the mean time between 
galaxy mergers.
Here we consider whether black hole coalescence can alter the spin 
axis of the larger black hole and yield a detectable geometric
signature in the radio observations of merging galaxies.

We estimate the effect of binary black hole coalescence on 
the spin of the resulting black hole using angular momentum conservation,
$$
\bf{S}_1 + \bf{S}_2 + \bf{L}_{\rm{orb}} = \bf{S} + \bf{J}_{\rm{rad}}
\eqno(1)
$$
where $\bf{S}_1$ and $\bf{S}_2$ are the spin angular momenta of the
two SBHs just before the final plunge,
$\bf{L}_{\rm orb}$ is the orbital angular momentum of the binary
before the plunge,
$\bf{S}$ is the spin of the resulting SBH, 
and $\bf{J}_{\rm{rad}}$ is the angular
momentum carried away by the gravitational waves during and after the coalescence.
Splitting of the angular momentum as in Eq. 1 is strictly only
defined in a post-Newtonian limit ({\it 3}) but this ambiguity is 
unimportant in what follows.

We denote the masses of the two SBHs as $M_1$ and $M_2$, 
such that $M_1\ge M_2$, and define their sum as $M\equiv M_1+M_2$.
Consider first the effect of the orbital angular momentum
on the spin of the resulting black hole.
The appropriate value for $L_{\rm orb}$ is the angular momentum
of the innermost stable circular orbit (ISCO) of the binary.
For $M_2\ll M_1$, this varies from $\sqrt{12} GM_1M_2/c$ if the larger
SBH is nonrotating, to $GM_1M_2/c$ for a prograde orbit around a
black hole that spins at the maximum possible rate, $S_1=GM_1^2/c$ ({\it 4}).
When $M_1\approx M_2$, the ISCO is not easily determined but various
approximations have been derived based on post-Newtonian expansions
and numerical calculations.

We estimated the contribution of $L_{\rm orb}$ to
the resultant spin in two ways.
Adding mass {\it gradually} to an initially nonrotating black hole
from a fixed plane produces
a spinup that depends uniquely on the total accreted mass $\delta M$; 
in this limiting case, gravitational radiation losses are 
negligible and the radius of the ISCO is known precisely ({\it 4}).
Alternatively, we can relate $S$ to $L_{\rm orb}$ using the approximate 
expression ({\it 5}) for the orbital angular 
momentum at ISCO in unequal-mass black hole binaries; 
this expression also ignores the gravitational radiation reaction.
The two approximations yield similar results (Fig. 1) suggesting that 
the spinup of an initially nonrotating black hole will exceed 
$\sim 1/2$ of the maximal spin $S_{\rm max}\equiv GM^2/c$ if the 
accreted mass exceeds just $\sim 1/5$ that of the primary.

\begin{figure}
\begin{center}
\includegraphics[width=.7\textwidth]{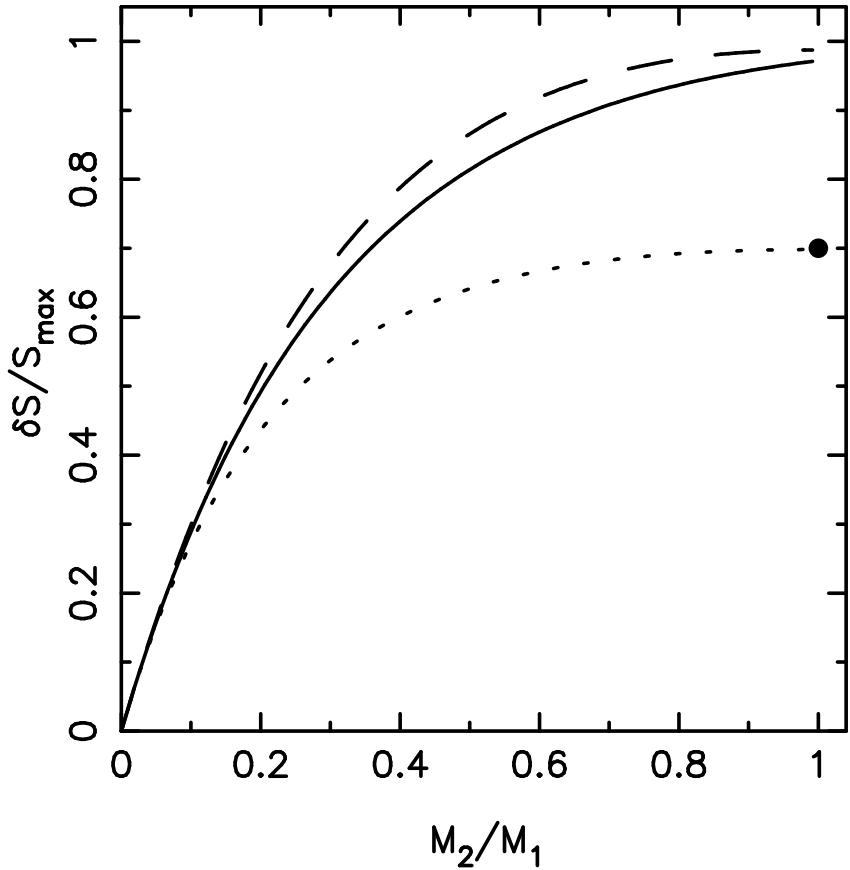}
\end{center}
\caption{
Spin angular momentum imparted to an initially nonrotating
black hole of mass $M_1$ by accretion of a mass $M_2$.
Accretion is assumed to take place from the innermost stable circular 
orbit (ISCO); $S_{\rm max}=G(M_1+M_2)^2/c$, the maximum allowed
angular momentum of the resulting hole.
Solid line: spinup produced by the gradual accretion of mass $M_2$
(ref. 4).
Dashed line: spinup resulting from coalescence with a second hole of
mass $M_2$ (ref. 5).
Both of these curves ignore gravitational radiation losses.
The filled circle is from the fully general-relativistic calculation
of Baker et al. (6).
The dotted line is a heuristic expression of Wilson \& Colbert (7)
that accounts for gravitational radiation losses,
normalized to go through the Baker et al. point.
}
\end{figure}

Loss of angular momentum by gravitational waves, $\bf{J}_{\rm rad}$,
becomes increasingly important as $M_2$ approaches $M_1$.
The fully general-relativistic calculations necessary for 
computing $\bf{J}_{\rm rad}$ have so far only been carried out for the case 
of an equal-mass, circular-orbit binary with no initial spins.
Baker et al. ({\it 6}) found that about 12\% of the system's total 
angular momentum was carried away by gravitational waves;
the final spin was $\sim 0.7 S_{\rm max}$ (Fig. 1).

We next consider changes in the orientation of the larger black hole's spin axis
due to the coalescence.
Let $\psi$ be the reorientation angle, i.e. the angle between the 
initial spin $\bf{S}_1$ of the more massive hole and the final 
spin $\bf{S}$.
Define $\lambda\equiv\delta S/S_1$, where 
$\delta\bf{S}\equiv\bf{S}-\bf{S}_1=\bf{L}_{\rm orb} + \bf{S}_2-\bf{J}_{\rm rad}$.
Then
$$
\cos\psi = {1+\lambda\cos\theta\over\sqrt{1+\lambda^2+2\lambda\cos\theta}}
\eqno(2)
$$
with $\theta$ the angle between $\delta\bf{S}$
and $\bf{S_1}$.
For $\lambda<1$ the maximum reorientation angle is 
$\cos\psi_{\rm max}=\sqrt{1-\lambda^2}$ while for $\lambda>1$
any value of $\psi$ is allowed.
Assuming a random orientation of $\delta\bf{S}$ with respect to
$\bf{S}_1$ (not strictly justified when $S_1>0$, because the radius of the ISCO
depends on the relative orientation of $\bf{L}_{\rm orb}$ and 
$\bf{S}_1$),
the distribution of $\psi$ can be computed and
the mean value of $\mu\equiv\cos\psi$ is
$$
\langle\mu\rangle=\cases {1-{\lambda^2\over 3},&$\lambda\le 1$;\cr
2\over 3\lambda,&$\lambda>1$.\cr}
\eqno(3)
$$

Comparing Eq. 3 with Fig. 1, we conclude that the reorientation angle 
is expected to be large when $M_2\gap 0.2 M_1$, even if the larger hole is
rapidly rotating initially.
For instance, accretion of a black hole with $M_2=M_1/4$ results in 
$\delta S\approx GM_1^2/2c$, comparable to $S_1$ if the larger 
black hole's spin is initially $\sim 1/2$ of its maximum value.
Hence $\lambda\gap 1$ and the reorientation angle could be as 
large as $180^{\circ}$, with a likely value of $\sim 50^{\circ}$.
If the larger black hole is slowly rotating initially, $S_1\ll GM_1^2/c$,
even smaller infalling black holes could produce substantial
reorientations; for instance, for $S_1=0.1GM_1^2/c$,
accretion of a black hole with $M_2\approx 0.05M_1$ 
can produce arbitrarily large realignments.

Although the reorientation of a SBH's spin axis due to coalescence
is not directly observable,
any gaseous accretion onto the SBH is constrained by 
relativistic frame dragging to be axisymmetric with respect to the
black hole ({\it 8}), and it is widely believed that the jets emitted 
from the centers of active galaxies are launched perpendicularly
to the inner accretion disk; hence a jet should point in the same 
direction as the spin axis of the SBH at its center ({\it 9}).
The extraordinary long-term stability of the jet direction in many 
radio galaxies is strong evidence that jet orientations are regulated 
by black hole spins ({\it 9,10}).
Because powerful radio galaxies comprise only a fraction, of order $1\%$, 
of all bright elliptical galaxies and because radio power is a rapidly 
increasing function of galaxy luminosity ({\it 11}), 
we would expect only the more massive of the two merging galaxies to 
harbor a jet.
Hence a likely consequence of SBH coalescence in a radio galaxy
is a sudden change in the direction of the jet associated 
with the larger SBH, followed by the generation of a new radio lobe
at some (possibly large) angle with respect to the original lobe.

In fact there is a class of radio sources which fit this
description: the so-called ``winged'' or ``X-type'' radio sources.
X-shaped sources are characterized by two low-surface-brightness radio 
lobes (the ``wings'') oriented at an angle to the ``active,'' 
or high-surface-brightness, lobes (Fig. 2); 
both sets of lobes pass symmetrically through the
center of the associated elliptical galaxy.
The first winged source discovered, NGC 326, was initially interpreted by
a model in which a single SBH undergoes slow geodetic precession
due to torques from an external mass, resulting in an S-shaped radio 
morphology ({\it 12}); 
later observations ({\it 13}) revealed the X shape
of this source indicating a more rapid change of jet direction.
Other explanations for the origin of X-shaped sources have been
proposed but none has proved satisfactory;
black holes are nearly perfect gyroscopes and reorienting them
via external forces is difficult.
One proposed model is based on a warping instability of accretion
disks ({\it 14}), but this model fails to explain why jet reorientation
occurs only once in the X-shaped sources and why most radio galaxies have
stable jet directions.
Capture of a dwarf galaxy with mass comparable to $M_1$ could
reorient a black hole but it is more likely that the infalling
galaxy would be disrupted by tidal forces before being accreted ({\it 15}).

\begin{figure}
\begin{center}
\includegraphics[width=.7\textwidth,bb= 150 250 438 538]{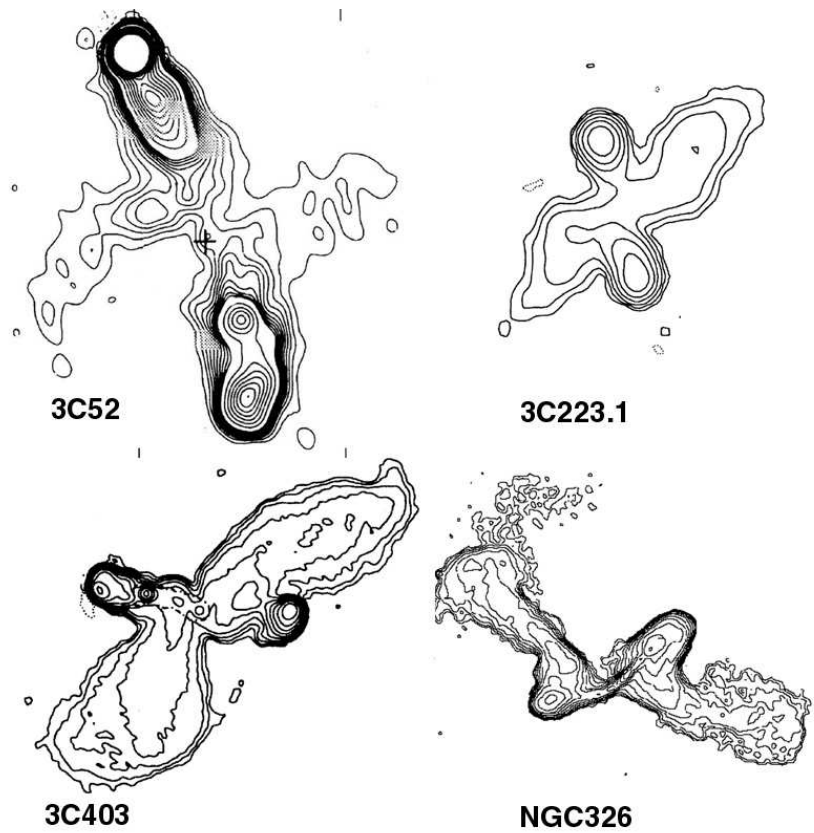}
\end{center}
\caption{
A composite of four clear examples of the X-shape morphology.
VLA radio observations of 3C52 (28),
3C223.1 (29), 3C403 (29),
and NGC 326 (13).
}
\end{figure}

Seven out of eleven X-shaped radio galaxies in which the length of
the wings is at least 80\% of the length of the active lobes have
Fanaroff-Riley type II (FRII); the others are either FRI 
(NGC 326) or mixed ({\it 16,17}; Table 1).
It has been argued that the host galaxies of FRII sources are
the products of recent mergers ({\it 7}).  
Because a major merger of comparably massive galaxies would presumably
have induced wiggles in the original jet due to bulk motion of the
galaxies, the fact that the wings are reasonably straight in 
most of these sources suggests that the mergers were minor ({\it 18}).
This is reasonable because minor mergers far outnumber major mergers,
and because the infalling SBH (whose mass should scale
roughly with host galaxy mass, ref. {\it 19}) need only have a 
fraction of the mass of the larger SBH in order to realign it.

The distinctness of the wings and active lobes in the X-shaped
galaxies suggests that
jet reorientation took place in a relatively short time,
$\lap 10^7$ yr ({\it 20}).
In our model, the reorientation of the SBHs occurs almost
instantaneously; most of the gravitational radiation accompanying the
coalescence of two, $10^8M_{\odot}$ SBHs would be emitted within just 
$\sim 100$ s ({\it 6}).
However the reorientation of the lobe-producing jets would take place 
on the longer time scale associated with Lense-Thirring precession of 
the inner accretion disk, 
$t_{\rm precess}\sim t_{\rm orb}(S_{max}/S)(r/r_s)^{3/2}$, where
$t_{orb}$ is the orbital period at radius $r$ and $r_s=2GM/c^2$
({\it 7}).
The time for reorientation is $\sim 1$ yr for the inner accretion disk,
and could be even longer if the continued powering of the jet required 
a realignment of gas at larger radii.
In addition, it may be a long time after the galactic merger before
the second black hole finds its way to the center of the merged
system and forms a binary which can coalesce.
The time scale for infall of the smaller galaxy and its SBH is 
$\sim 2\times 10^8 {\rm yr\ }\sigma^5_{200}M^{-3/4}_{2,7}$
where $\sigma_{200}$ is the velocity dispersion of the larger galaxy
in units of $200$ km s$^{-1}$ and 
$M_{2,7}=M_2/10^7 M_{\odot}$ ({\it 21}).
Hence we would not necessarily expect to see signs of a recent
merger in the morphology of the radio source's host galaxy.

We speculate that the time scale for accretion disk realignment
may influence the radio source morphology.
Slow realignment would cause the jet to deposit its
energy into a large volume of space, leading to an FRI
source with S-type morphology.
Rapid realignment would produce an intermediate-luminosity
X-shaped source, perhaps with a radio power near the FRI/FRII break.
If realignment occurred long ago ($\gap 10^8$ yr), the jets and lobes 
would be well aligned and the source could build up to a high-luminosity
FRII source.

The probability of observing a radio galaxy as an X-shaped
source in our model is $\sim T_X/T_{\rm merge}$,
where $T_X$ is the length of time that the wings remain visible
and $T_{\rm merge}$ is the mean time between mergers.
The fading time $T_X$ can be estimated in a number of ways.
Table 1 lists estimates of the time since the radiating particles
in the wings were last accelerated;
if the mean age of the X-sources is of order 
their visible lifetime, the spectral aging estimates in Table 1 
imply $T_X\lap 10^8$ yr.
These spectral aging estimates are somewhat model dependent
due to the effects of particle re-acceleration, but we can
make a direct, semi-empirical estimate of the age
by noting the similarity between the wings in the X-sources
and the end of the plasma trail in the so-called
narrow angle tail sources found in groups and clusters.
Luminosity, spectral shape, polarization and brightness
are similar and so we would expect both types of
source to be visible for a similar time.
The galaxies generating these tail sources have velocities
typical of the cluster population -- we can estimate 
a dynamical age of $\sim 10^8$ yr where the tail fades
away ({\it 22}).

If we accept $T_X\approx 10^8$ yr, and use Leahy \& Parma's ({\it 16})
estimate that $\sim 7\%$ of radio galaxies in their sample are X-sources,
the mean merger rate for the radio galaxy sample
becomes $0.07/10^8 {\rm yr}\approx 1$ Gyr$^{-1}$.
This rate is higher than most estimates of the overall galaxy
merger rate but it should be a fair estimate unless
there is a correlation between the presence of the radio source
and the population of galaxies undergoing mergers.
There are indeed reasons to believe that there may be
such a correlation ({\it 23}) and more speculatively the 
black hole coalescence itself may be the trigger for the active
galaxy phenomenon.
Any such correlation would decrease the implied merger rate for the
galaxy population as a whole but rates of $\sim 1$ Gyr$^{-1}$ are 
typical of those inferred for galaxies in dense regions or 
groups ({\it 24}).
Our result should motivate more detailed studies of galaxy
mergers in the hope of demonstrating that binary SBHs can
indeed avoid ``stalling'' and go on to rapid coalescence.

If the coalescence rate of binary SBHs is comparable 
to the galaxy merger rate, then the binary
separation must be able to drop from $\sim 1$ pc to $\sim 0.01$ pc in a time
shorter than $\sim 1$ Gyr.
The predicted event rate for gravitational wave interferometers should 
then be about equal to the integrated galaxy merger rate
out to a redshift $z\approx 5$,
implying a time between detections of $\sim 1$ yr ({\it 25}).

\clearpage

{\bf References and Notes}

\begin{enumerate}

\item P. L. Bender,  
in {\it Gravitational Waves} 
(ed. Ciufolini, I. et al.) 115
(Institute of Physics Publishing, Bristol, UK, 2001).

\item M. Milosavljevic, D. Merritt, 
{\it Astrophys. J.}
{\bf 563}, 34 (2001).

\item E. E. Flanagan, S. A. Hughes, 
{\it Phys. Rev. D} {\bf 57}, 4535
(1998).

\item J. M. Bardeen, 
{\it Nature} {\bf 226}, 64 (1970).

\item L. E. Kidder, C. M. Will, A. G. Wiseman,
{\it Phys. Rev. D} 
{\bf 47}, 3281 (1993).

\item J. Baker, M. Campanelli, C. O. Lousto, R. Takahashi,
astro-ph/0202469 (2002).

\item A. S. Wilson, E. J. M. Colbert, 
{\it Astrophys. J.} {\bf 438}, 62 (1995).

\item J. M. Bardeen, J. A. Petterson,
{\it Astrophys. J.} {\bf 195}, L65 (1975).

\item M. J. Rees, 
{\it Nature} {\bf 275}, 516 (1978).

\item M. C. Begelman, R. D. Blandford, M. J. Rees,
{\it Rev. Mod. Phys.}
{\bf 56}, 255 (1984).

\item C. Auriemma et al. 
{\it Astron. Astrophys.} {\bf 57}, 41 (1977).

\item R. D. Ekers, R. Fanti, C. Lari, P. Parma,
{\it Nature} {\bf 276}, 588 (1978).

\item M. Murgia, P. Parma, H. R. Ruiter, M. Bondi,
Ekers, R. D., Fanti, R. \& Fomalont, E. B. 
{\it Astron. Astrophys.} {\bf 380}, 102 (2001).

\item J. E. Pringle, 
{\it Mon. Not. R. Astron. S.} {\bf 281}, 357 (1996).

\item D. Merritt \& F. Cruz,
{\it Astrophys. J.} {\bf 551}, L4a (2001).

\item J. P. Leahy, P. Parma,
In {\it Extragalactic Radio Sources. From Beams to Jets.}
(ed. Roland, J. et al.) 307 (Cambridge University Press,
Cambridge, England, 1992).

\item The Fanaroff-Riley types I and II are a morphological
classification defined by the degree of edge-brightening in
the radio lobes; see B. Fanaroff, J. Riley,
{\it Mon. Not. R. Astron. S.} {\bf 167}, 31P (1974).

\item A minor merger is one in which the infalling galaxy
has a mass that is 30\% or less that of the larger galaxy.

\item D. Merritt, L. Ferrarese,
{\it Mon. Not. R. Astron. Soc.} {\bf 320}, L30 (2001).

\item J. Dennett-Thorpe et al.
Jet reorientation in AGN: Two-winged radio galaxies.
{\it Mon. Not. R. Astron. Soc.} {\bf 330}, 609 (2002).

\item Merritt, D.
In {\it XVth IAP Meeting Dynamics
of Galaxies: From the Early Universe to the Present} (eds. Combes, F.
et al.) Vol. 197, 221
(ASP Conf. Ser., Astronomical Society of the Pacific, San Francisco, 2000) 

\item C. P. O'Dea, R. N. Owen, 
{\it Astron. J.} {\bf 90}, 954 (1985).

\item T. M. Heckman et al.
{\it Astrophys. J.} {\bf 311}, 526 (1986).

\item R. G. Carlberg et al.
{\it Astrophys. J.} {\bf 532}, L1 (2000).

\item M. G. Haehnelt, 
In {\it Laser Interferometer Space Antenna} (ed. Folkner, W. M.) 
Vol. 456, 45 (AIP Conference Proceedings, American Institute of Physics,
Woodbury, NY, 1998)

\item M. Murgia, 
PhD Thesis, University of Bologna (2000).

\item U. Klein, K.-H. Mack L. Gregorini, P. Parma,
{\it Astron. Astrophys.} {\bf 303},
427 (1995).

\item J. P. Leahy, A. G. Williams,
{\it Mon. Not. R. Astron. Soc.} {\bf 210}, 929 (1984)

\item J. Dennett-Thorpe, A. H. Bridle, R. A. Laing, 
P. A. G. Scheuer, 
{\it Mon. Not. R. Astron. Soc.} {\bf 304}, 27 (1999).

\item P. Leahy kindly allowed us to reproduce the Leahy \& Parma
catalog of X-shaped radio sources (ref. 16) in Table 1.
We thank him, L. Ferrarese, S. Hughes, M. Milosavljevic, and C. O'Dea 
for useful discussions.
This work was supported by the National Science 
Foundation through grant 4-21911, and by the National Aeronautics and
Space Administration through grants 4-21904 and NAG5-8693,
and by the Miller Institute, University of California at Berkeley.

\end{enumerate}

\clearpage

\noindent
Table 1. X-shaped radio sources from the compilation of
Leahy \& Parma (15). $z$ is redshift;
$m$ is total visual magnitude of the host galaxy;
$T$ is estimated time since the relativistic electrons
were last accelerated in the wings of the sources.

$$\vbox { \halign {#\hfil & # & \hfil #\hfil & # & \hfil # \hfil & # & \hfil#\hfil & # & #\hfil \cr
Name & & z & & m & & $T$ & & Notes \cr
3C52         & & 0.2854 & & 18.5 & &  & & Dust disk      \cr
3C136.1      & & 0.0640 & & 17   & &  & & Double nucleus/merger remnant \cr
3C223.1      & & 0.1075 & & 16.6 & &$<35$ Myr ({\it 20})&&Dust disk\cr
3C315        & & 0.1083 & & 18.3 & &           & & Substructure      \cr
3C403        & & 0.0590 & & 16.5 & &$<17$ Myr  ({\it 20}) & &\cr
3C433        & & 0.1016 & &      & &           & & Dust/star formation\cr
4C12.03      & & 0.1100 & & 17.8 & &           & &                \cr
4C48.29      & & 0.0530 & & 16.0 & &           & &                \cr
B2 0055+26   & & 0.0487 & & 13.0 & &$\sim70$ Myr ({\it 26})&&Double galaxy  \cr
(NGC 326)    & &        & &      & &           & &                \cr
B2 0828+32   & & 0.0527 & & 15.1 & &$<75$ Myr ({\it 27}) & & \cr
B1059+169    & & 0.0677 & & 15.2 & &           & &                \cr
(Abell 1145) & &        & &      & &           & &                \cr\cr
\cr
}}
$$

\end{document}